\documentclass[screen=true, sigplan]{acmart}
\usepackage[utf8]{inputenc}
\usepackage[T1]{fontenc}

\newcommand{\packageGraphicx}{\usepackage{graphicx}}
\newcommand{\packageHyperref}{\usepackage{hyperref}}
\newcommand{\renewrmdefault}{\renewcommand{\rmdefault}{ptm}}
\newcommand{\packageRelsize}{\usepackage{relsize}}
\newcommand{\packageAmsmath}{\usepackage{amsmath}}
\newcommand{\packageMathabx}{\usepackage{mathabx}}
\newcommand{\packageWasysym}{
  \let\leftmoon\relax \let\rightmoon\relax \let\fullmoon\relax \let\newmoon\relax \let\diameter\relax
  \usepackage[nointegrals]{wasysym}}
\newcommand{\packageTxfonts}{
  \let\widering\relax
  \let\oldwidebar\widebar
  \let\widebar\relax
  \usepackage{newtxmath}
  \ifx\widebar\relax
    \let\widebar\oldwidebar
  \fi
}
\newcommand{\packageTextcomp}{\usepackage{textcomp}}
\newcommand{\packageFramed}{\usepackage{framed}}
\newcommand{\packageHyphenat}{\usepackage[htt]{hyphenat}}
\newcommand{\packageColor}{\usepackage[usenames,dvipsnames]{color}}
\newcommand{\doHypersetup}{\hypersetup{bookmarks=true,bookmarksopen=true,bookmarksnumbered=true}}
\newcommand{\packageTocstyle}{}
\newcommand{\packageCJK}{\IfFileExists{CJK.sty}{\usepackage{CJK}}{}}
\renewcommand\packageColor\relax
\renewcommand\packageTocstyle\relax
\renewcommand\packageMathabx{\ifx\bigtimes\undefined \usepackage{mathabx} \else \relax \fi}
\renewcommand\packageTxfonts\relax

\renewcommand{\renewrmdefault}{}

\packageGraphicx
\packageHyperref
\renewrmdefault
\packageRelsize
\packageAmsmath
\packageMathabx
\packageWasysym
\packageTxfonts
\packageTextcomp
\packageFramed
\packageHyphenat
\packageColor
\doHypersetup
\packageTocstyle
\packageCJK

\newcommand{\sectionNewpage}{}

\newcommand{\preDoc}{}
\newcommand{\postDoc}{}

\newcommand{\BookRef}[2]{\emph{#2}}
\newcommand{\ChapRef}[2]{\SecRef{#1}{#2}}
\newcommand{\SecRef}[2]{section~#1}
\newcommand{\PartRef}[2]{part~#1}
\newcommand{\BookRefUC}[2]{\BookRef{#1}{#2}}
\newcommand{\ChapRefUC}[2]{\SecRefUC{#1}{#2}}
\newcommand{\SecRefUC}[2]{Section~#1}
\newcommand{\PartRefUC}[2]{Part~#1}

\newcommand{\BookRefLocal}[3]{\hyperref[#1]{\BookRef{#2}{#3}}}
\newcommand{\ChapRefLocal}[3]{\hyperref[#1]{\ChapRef{#2}{#3}}}
\newcommand{\SecRefLocal}[3]{\hyperref[#1]{\SecRef{#2}{#3}}}
\newcommand{\PartRefLocal}[3]{\hyperref[#1]{\PartRef{#2}{#3}}}
\newcommand{\BookRefLocalUC}[3]{\hyperref[#1]{\BookRefUC{#2}{#3}}}
\newcommand{\ChapRefLocalUC}[3]{\hyperref[#1]{\ChapRefUC{#2}{#3}}}
\newcommand{\SecRefLocalUC}[3]{\hyperref[#1]{\SecRefUC{#2}{#3}}}
\newcommand{\PartRefLocalUC}[3]{\hyperref[#1]{\PartRefUC{#2}{#3}}}

\newcommand{\BookRefUN}[1]{\BookRef{}{#1}}

\newcommand{\SecRefUN}[1]{``#1''}

\newcommand{\BookRefLocalUN}[2]{\hyperref[#1]{\BookRefUN{#2}}}

\newcommand{\SecRefLocalUN}[2]{\hyperref[#1]{\SecRefUN{#2}}}

\newcommand{\SectionNumberLink}[2]{\hyperref[#1]{#2}}

\newcommand{\textsuper}[1]{$^{\hbox{\textsmaller{#1}}}$}

\newcommand{\planetName}[1]{PLane\hspace{-0.1ex}T}

\makeatletter
\def\empty@finalstrut#1{%
  \unskip\ifhmode\nobreak\fi\vrule\@width\z@\@height\z@\@depth\z@}
\def\no@strut{\global\setbox\@arstrutbox\hbox{%
    \vrule \@height\z@
           \@depth\z@
           \@width\z@}%
    \gdef\@endpbox{\empty@finalstrut\@arstrutbox\par\egroup\hfil}%
}%
\def\yes@strut{\global\setbox\@arstrutbox\hbox{%
    \vrule \@height\arraystretch \ht\strutbox
           \@depth\arraystretch \dp\strutbox
           \@width\z@}%
    \gdef\@endpbox{\@finalstrut\@arstrutbox\par\egroup\hfil}%
}%
\def\@mkpream#1{\@firstamptrue\@lastchclass6
  \let\@preamble\@empty\def\empty@preamble{\add@ins}%
  \let\protect\@unexpandable@protect
  \let\@sharp\relax\let\add@ins\relax
  \let\@startpbox\relax\let\@endpbox\relax
  \@expast{#1}%
  \expandafter\@tfor \expandafter
    \@nextchar \expandafter:\expandafter=\reserved@a\do
       {\@testpach\@nextchar
    \ifcase \@chclass \@classz \or \@classi \or \@classii \or \@classiii
      \or \@classiv \or\@classv \fi\@lastchclass\@chclass}%
  \ifcase \@lastchclass \@acol
      \or \or \@preamerr \@ne\or \@preamerr \tw@\or \or \@acol \fi}
\def\@addamp{%
  \if@firstamp
    \@firstampfalse
    \edef\empty@preamble{\add@ins}%
  \else
    \edef\@preamble{\@preamble &}%
    \edef\empty@preamble{\expandafter\noexpand\empty@preamble &\add@ins}%
  \fi}
\newif\iftw@hlines \tw@hlinesfalse
\def\@xhline{\ifx\reserved@a\hline
               \tw@hlinestrue
             \else\ifx\reserved@a\Hline
               \tw@hlinestrue
             \else
               \tw@hlinesfalse
             \fi\fi
      \iftw@hlines
        \aftergroup\do@after
      \fi
      \ifnum0=`{\fi}%
}
\def\do@after{\emptyrow[\the\doublerulesep]}
\def\emptyrow{\noalign\bgroup\@ifnextchar[\@emptyrow{\@emptyrow[\z@]}}
\def\@emptyrow[#1]{\no@strut\gdef\add@ins{\vrule \@height\z@ \@depth#1 \@width\z@}\egroup%
\empty@preamble\\
\noalign{\yes@strut\gdef\add@ins{\vrule \@height\z@ \@depth\z@ \@width\z@}}%
}
\def\tabrow#1{\noalign\bgroup\@ifnextchar[{\@tabrow{#1}}{\@tabrow{#1}[]}}
\def\@tabrow#1[#2]{\no@strut\egroup#1\ifx.#2.\\\else\\[#2]\fi\noalign{\yes@strut}}
\def\endpltstabular{\crcr\egroup\egroup \egroup}
\expandafter \let \csname endpltstabular*\endcsname = \endpltstabular
\def\pltstabular{\let\@halignto\@empty\@pltstabular}
\@namedef{pltstabular*}#1{\def\@halignto{to#1}\@pltstabular}
\def\@pltstabular{\leavevmode \bgroup \let\@acol\@tabacol
   \let\@classz\@tabclassz
   \let\@classiv\@tabclassiv \let\\\@tabularcr\@stabarray}
\def\@stabarray{\m@th\@ifnextchar[\@sarray{\@sarray[c]}}
\def\@sarray[#1]#2{%
  \bgroup
  \setbox\@arstrutbox\hbox{%
    \vrule \@height\arraystretch\ht\strutbox
           \@depth\arraystretch \dp\strutbox
           \@width\z@}%
  \@mkpream{#2}%
  \edef\@preamble{%
    \ialign \noexpand\@halignto
      \bgroup \@arstrut \@preamble \tabskip\z@skip \cr}%
  \let\@startpbox\@@startpbox \let\@endpbox\@@endpbox
  \let\tabularnewline\\%
    \let\@sharp##%
    \set@typeset@protect
    \lineskip\z@skip\baselineskip\z@skip
    \@preamble}

\makeatother

\newlength{\stabLeft}

\newenvironment{SingleColumn}{\begin{list}{}{\topsep=0pt\partopsep=0pt%
\listparindent=0pt\itemindent=0pt\labelwidth=0pt\leftmargin=0pt\rightmargin=0pt%
\itemsep=0pt\parsep=0pt}\item}{\end{list}}

\newcommand{\Sendabbrev}[1]{#1\@}

\newcommand{\SCodePreSkip}{\vskip\abovedisplayskip}
\newcommand{\SCodePostSkip}{\vskip\belowdisplayskip}

\newcommand{\SVInsetPreSkip}{\vskip\abovedisplayskip}
\newcommand{\SVInsetPostSkip}{\vskip\belowdisplayskip}

\newcommand{\titleAndVersionAndAuthors}[3]{\title{#1\\{\normalsize \SVersionBefore{}#2}}\author{#3}\maketitle}

\newcommand{\titleAndEmptyVersionAndAuthors}[3]{\title{#1}\author{#3}\maketitle}

\newcommand{\SAuthor}[1]{#1}
\newcommand{\SAuthorSep}[1]{\qquad}
\newcommand{\SVersionBefore}[1]{Version }

\newcommand{\SNumberOfAuthors}[1]{}

\let\SOriginalthesubsection\thesubsection
\let\SOriginalthesubsubsection\thesubsubsection

\newcommand{\Ssection}[2]{\section[#1]{#2}\let\thesubsection\SOriginalthesubsection}
\newcommand{\Ssubsection}[2]{\subsection[#1]{#2}\let\thesubsubsection\SOriginalthesubsubsection}

\newcommand{\Ssectionstar}[1]{\section*{#1}\renewcommand*\thesubsection{\arabic{subsection}}\setcounter{subsection}{0}}

\newcommand{\Ssubsubsubsectionstar}[1]{{\bf #1}}
\newcommand{\Ssubsubsubsubsectionstar}[1]{\Ssubsubsubsectionstar{#1}}

\newcommand{\Ssectionstarx}[2]{\Ssectionstar{#2}\phantomsection\addcontentsline{toc}{section}{#1}}

\newcounter{GrouperTemp}

\newcommand{\SSubSubSubSection}[1]{\Ssubsubsubsubsectionstar{#1}}

\newcommand{\Snolinkurl}[1]{\nolinkurl{#1}}

\newcommand{\SAuthorinfo}[4]{#1}
\newcommand{\SAuthorPlace}[1]{#1}
\newcommand{\SAuthorEmail}[1]{#1}
\newcommand{\SAuthorOrcid}[1]{#1}

\newcommand{\SConferenceInfo}[2]{}
\newcommand{\SCopyrightYear}[1]{}
\newcommand{\SCopyrightData}[1]{}
\newcommand{\Sdoi}[1]{}

\newcommand{\SCategory}[3]{}
\newcommand{\SCategoryPlus}[4]{}
\newcommand{\STerms}[1]{}
\newcommand{\SKeywords}[1]{}

\newcommand{\AutobibLink}[1]{\color{ACMPurple}{#1}}

\newenvironment{AutoBibliography}{\begin{small}}{\end{small}}
\newcommand{\Autobibentry}[1]{\hspace{0.05\linewidth}\parbox[t]{0.95\linewidth}{\parindent=-0.05\linewidth#1\vspace{1.0ex}}}
\newcommand{\Autobibtarget}[1]{\phantomsection#1}

\usepackage{calc}
\newlength{\ABcollength}

\newcommand{\Autobibref}[1]{#1}

\providecommand{\AutobibLink}[1]{#1}

\newcommand{\pseudodoi}[1]{#1}

\newcommand{\NoteBox}[1]{\footnote{#1}}
\newcommand{\NoteContent}[1]{#1}

\newcommand{\FootnoteRef}[1]{}
\newcommand{\FootnoteTarget}[1]{}

\newcommand{\FootnoteBlockContent}[1]{}
\usepackage{ccaption}

\makeatletter
\@ifundefined{belowcaptionskip}{}{}
\makeatother

\newcommand{\Legend}[1]{~

                        \hrule width \hsize height .33pt
                        \vspace{4pt}
                        \legend{#1}}

\newcommand{\FigureTarget}[2]{#1}
\newcommand{\FigureRef}[2]{#1}

\newlength{\FigOrigskip}
\FigOrigskip=\parskip

\newcommand{\FigureSetRef}{\refstepcounter{figure}}

\newenvironment{Figure}{\begin{figure}\FigureSetRef}{\end{figure}}
\newenvironment{FigureMulti}{\begin{figure*}[t!p]\FigureSetRef}{\end{figure*}}

\newenvironment{Centerfigure}{\begin{Xfigure}\centering\item}{\end{Xfigure}}

\newenvironment{Xfigure}{\begin{list}{}{\leftmargin=0pt\topsep=0pt\parsep=\FigOrigskip\partopsep=0pt}}{\end{list}}

\newenvironment{FigureInside}{}{}

\newcommand{\Centertext}[1]{\begin{center}#1\end{center}}

\usepackage{savesym}
\savesymbol{iint}
\savesymbol{iiint}
\savesymbol{dddot}
\savesymbol{ddddot}
\savesymbol{overleftrightarrow}
\savesymbol{underrightarrow}
\savesymbol{underleftarrow}
\savesymbol{underleftrightarrow}
\usepackage{amsmath}
\restoresymbol{AMS}{iint}
\restoresymbol{AMS}{iiint}
\restoresymbol{AMS}{dddot}
\restoresymbol{AMS}{ddddot}
\restoresymbol{AMS}{overleftrightarrow}
\restoresymbol{AMS}{underrightarrow}
\restoresymbol{AMS}{underleftarrow}
\restoresymbol{AMS}{underleftrightarrow}
\savesymbol{ulcorner}
\savesymbol{urcorner}
\savesymbol{llcorner}
\savesymbol{lrcorner}
\usepackage{amsfonts}
\restoresymbol{AMS}{ulcorner}
\restoresymbol{AMS}{urcorner}
\restoresymbol{AMS}{llcorner}
\restoresymbol{AMS}{lrcorner}
\newcommand{\Iidentity}[1]{#1}

\usepackage{amsthm}
\newtheorem{definition}{Definition}
\newtheorem{theorem}{Theorem}
\newtheorem{conjecture}{Conjecture}

\renewcommand{\titleAndVersionAndAuthors}[3]{\title{#1}#3\maketitle}
\renewcommand{\titleAndEmptyVersionAndAuthors}[3]{\titleAndVersionAndAuthors{#1}{#2}{#3}}

\def\SAuthor#1{\SAutoAuthor#1\SAutoAuthorDone{#1}}
\def\SAutoAuthorDone#1{}
\def\SAutoAuthor{\futurelet\next\SAutoAuthorX}
\def\SAutoAuthorX{\ifx\next\SAuthorinfo \let\Snext\relax \else \let\Snext\SToAuthorDone \fi \Snext}
\def\SToAuthorDone{\futurelet\next\SToAuthorDoneX}
\def\SToAuthorDoneX#1{\ifx\next\SAutoAuthorDone \let\Snext\SAddAuthorInfo \else \let\Snext\SToAuthorDone \fi \Snext}
\newcommand{\SAddAuthorInfo}[1]{\SAuthorinfo{#1}{}{}}

\renewcommand{\SAuthorinfo}[4]{\author{#1}{#2}{#3}{#4}}
\renewcommand{\SAuthorSep}[1]{}
\renewcommand{\SAuthorOrcid}[1]{\orcid{#1}}
\renewcommand{\SAuthorPlace}[1]{\affiliation{#1}}
\renewcommand{\SAuthorEmail}[1]{\email{#1}}

\renewcommand{\SConferenceInfo}[2]{\conferenceinfo{#1}{#2}}
\renewcommand{\SCopyrightYear}[1]{\copyrightyear{#1}}
\renewcommand{\SCopyrightData}[1]{\copyrightdata{#1}}

\renewcommand{\SCategory}[3]{\category{#1}{#2}{#3}}
\renewcommand{\SCategoryPlus}[4]{\category{#1}{#2}{#3}[#4]}
\renewcommand{\STerms}[1]{\terms{#1}}
\renewcommand{\SKeywords}[1]{\keywords{#1}}

\usepackage{microtype}
\usepackage{doi}
\usepackage{letltxmacro}
\renewcommand{\Sdoi}[1]{\doi{#1}}

\LetLtxMacro{\oldfigure}{\Figure}
\LetLtxMacro{\oldendfigure}{\endFigure}

\renewenvironment{Figure}
{\oldfigure}
{\vspace{-2ex}\oldendfigure}

\setcopyright{none}
\acmDOI{}
\acmISBN{}
\acmPrice{}
\acmBooktitle{Workshop on Principles of Secure Compilation (PriSC)}
\acmConference[PriSC]{Workshop on Principles of Secure Compilation}{January 2021}{Online}
\begin{document}
\preDoc

\acmYear{2021}

\acmMonth{1}\titleAndEmptyVersionAndAuthors{Compilation as Multi{-}Language Semantics}{}{\SNumberOfAuthors{1}\SAuthor{\SAuthorinfo{William J. Bowman}{\SAuthorOrcid{0000-0002-6402-4840}}{\SAuthorPlace{\institution{University of British Columbia}\city{Vancouver}\state{BC}\country{CA}}}{\SAuthorEmail{wjb@williamjbowman.com}}}}
\label{t:x28part_x22Compilationx5fasx5fMultix2dLanguagex5fSemanticsx22x29}

\noindent 

\sectionNewpage

\Ssection{Extended Abstract}{Extended Abstract}\label{t:x28part_x22Extendedx5fAbstractx22x29}

Modeling interoperability between programs in different languages is a key
problem when modeling compositional and secure compilation.
Multi{-}language semantics provide a syntactic method for modeling language
interopability\Autobibref{~(\hyperref[t:x28autobib_x22Robert_Bruce_Matthews_Jacob_And_FindlerOperational_Semantics_for_Multix2dlanguage_ProgramsIn_Procx2e_Symposium_on_Principles_of_Programming_Languages_x28POPLx292007doix3a10x2e1145x2f1190216x2e1190220x22x29]{\AutobibLink{Matthews}} \hyperref[t:x28autobib_x22Robert_Bruce_Matthews_Jacob_And_FindlerOperational_Semantics_for_Multix2dlanguage_ProgramsIn_Procx2e_Symposium_on_Principles_of_Programming_Languages_x28POPLx292007doix3a10x2e1145x2f1190216x2e1190220x22x29]{\AutobibLink{2007}})}, and has proven useful in compiler
correctness and secure compilation\Autobibref{~(\hyperref[t:x28autobib_x22Amal_AhmedVerified_Compilers_for_a_Multix2dlanguage_WorldIn_Procx2e_Summit_oN_Advances_in_Programming_Languages_x28SNAPLx292015doix3a10x2e4230x2fLIPIcsx2eSNAPLx2e2015x2e15x22x29]{\AutobibLink{Ahmed}} \hyperref[t:x28autobib_x22Amal_AhmedVerified_Compilers_for_a_Multix2dlanguage_WorldIn_Procx2e_Summit_oN_Advances_in_Programming_Languages_x28SNAPLx292015doix3a10x2e4230x2fLIPIcsx2eSNAPLx2e2015x2e15x22x29]{\AutobibLink{2015}}; \hyperref[t:x28autobib_x22Amal_Ahmed_and_Matthias_BlumeAn_Equivalencex2dPreserving_CPS_Translation_via_Multix2dLanguage_SemanticsIn_Procx2e_International_Conference_on_Functional_Programming_x28ICFPx292011doix3a10x2e1145x2f2034773x2e2034830x22x29]{\AutobibLink{Ahmed and Blume}} \hyperref[t:x28autobib_x22Amal_Ahmed_and_Matthias_BlumeAn_Equivalencex2dPreserving_CPS_Translation_via_Multix2dLanguage_SemanticsIn_Procx2e_International_Conference_on_Functional_Programming_x28ICFPx292011doix3a10x2e1145x2f2034773x2e2034830x22x29]{\AutobibLink{2011}}; \hyperref[t:x28autobib_x22Max_Sx2e_Newx2c_William_Jx2e_Bowmanx2c_and_Amal_AhmedFully_Abstract_Compilation_via_Universal_EmbeddingIn_Procx2e_International_Conference_on_Functional_Programming_x28ICFPx292016doix3a10x2e1145x2f2951913x2e2951941x22x29]{\AutobibLink{New et al\Sendabbrev{.}}} \hyperref[t:x28autobib_x22Max_Sx2e_Newx2c_William_Jx2e_Bowmanx2c_and_Amal_AhmedFully_Abstract_Compilation_via_Universal_EmbeddingIn_Procx2e_International_Conference_on_Functional_Programming_x28ICFPx292016doix3a10x2e1145x2f2951913x2e2951941x22x29]{\AutobibLink{2016}}; \hyperref[t:x28autobib_x22Daniel_Patterson_and_Amal_AhmedLinking_Types_for_Multix2dLanguage_Softwarex3a_Have_Your_Cake_and_Eat_It_TooIn_Procx2e_Summit_oN_Advances_in_Programming_Languages_x28SNAPLx292017doix3a10x2e4230x2fLIPIcsx2eSNAPLx2e2017x2e12x22x29]{\AutobibLink{Patterson and Ahmed}} \hyperref[t:x28autobib_x22Daniel_Patterson_and_Amal_AhmedLinking_Types_for_Multix2dLanguage_Softwarex3a_Have_Your_Cake_and_Eat_It_TooIn_Procx2e_Summit_oN_Advances_in_Programming_Languages_x28SNAPLx292017doix3a10x2e4230x2fLIPIcsx2eSNAPLx2e2017x2e12x22x29]{\AutobibLink{2017}}; \hyperref[t:x28autobib_x22James_Tx2e_Perconti_and_Amal_AhmedVerifying_an_Open_Compiler_Using_Multix2dlanguage_SemanticsIn_Procx2e_European_Symposium_on_Programming_x28ESOPx292014doix3a10x2e1007x2f978x2d3x2d642x2d54833x2d8x5f8x22x29]{\AutobibLink{Perconti and Ahmed}} \hyperref[t:x28autobib_x22James_Tx2e_Perconti_and_Amal_AhmedVerifying_an_Open_Compiler_Using_Multix2dlanguage_SemanticsIn_Procx2e_European_Symposium_on_Programming_x28ESOPx292014doix3a10x2e1007x2f978x2d3x2d642x2d54833x2d8x5f8x22x29]{\AutobibLink{2014}})}.

Unfortunately, existing models of compilation using multi{-}language semantics
duplicate effort.
Two variants of each compiler pass are defined: a syntactic translation on open
terms, and a run{-}time translation of closed terms at multi{-}language
boundaries\Autobibref{~(\hyperref[t:x28autobib_x22Amal_Ahmed_and_Matthias_BlumeAn_Equivalencex2dPreserving_CPS_Translation_via_Multix2dLanguage_SemanticsIn_Procx2e_International_Conference_on_Functional_Programming_x28ICFPx292011doix3a10x2e1145x2f2034773x2e2034830x22x29]{\AutobibLink{Ahmed and Blume}} \hyperref[t:x28autobib_x22Amal_Ahmed_and_Matthias_BlumeAn_Equivalencex2dPreserving_CPS_Translation_via_Multix2dLanguage_SemanticsIn_Procx2e_International_Conference_on_Functional_Programming_x28ICFPx292011doix3a10x2e1145x2f2034773x2e2034830x22x29]{\AutobibLink{2011}}; \hyperref[t:x28autobib_x22Max_Sx2e_Newx2c_William_Jx2e_Bowmanx2c_and_Amal_AhmedFully_Abstract_Compilation_via_Universal_EmbeddingIn_Procx2e_International_Conference_on_Functional_Programming_x28ICFPx292016doix3a10x2e1145x2f2951913x2e2951941x22x29]{\AutobibLink{New et al\Sendabbrev{.}}} \hyperref[t:x28autobib_x22Max_Sx2e_Newx2c_William_Jx2e_Bowmanx2c_and_Amal_AhmedFully_Abstract_Compilation_via_Universal_EmbeddingIn_Procx2e_International_Conference_on_Functional_Programming_x28ICFPx292016doix3a10x2e1145x2f2951913x2e2951941x22x29]{\AutobibLink{2016}})}.
One must then prove that both definitions coincide.

We introduce a novel work{-}in{-}progress approach to uniformly model both variants
as a single reduction system on open terms in a multi{-}language semantics.
This simultaneously defines the compiler and the interoperability semantics.
It also has interesting semantic consequences: different reduction
strategies model different compilation strategies, and standard theorems about
reduction imply standard compiler correctness theorems.
For example, we get a model of ahead{-}of{-}time (AOT) compilation by normalizing
cross{-}language redexes; the normal form with respect to these redexes is a
target language term.
We model just{-}in{-}time (JIT) compilation as nondeterministic evaluation in the
multi{-}language: a term can either step in the source, or translate then step in
the target.
We prove that confluence of multi{-}language reduction implies compiler
correctness and part of full abstraction; and that subject reduction
implies type{-}preservation of the compiler.

\SSubSubSubSection{An example instance: Reduction to A{-}normal form}

Our approach generalizes from high{-}level to low{-}level transformations of a
wide array of language features.
To demonstrate this, we have developed a 5{-}pass model compiler from a
Scheme{-}like language to an x86{-}64{-}like language.
Here, we model one interesting compiler pass: reduction to A{-}normal form (ANF).
This pass is a good example and stress test.
The A{-}reductions are tricky to define because they reorder a term with respect
to its context, while the other passes locally transform a term in an arbitrary
context.

The source is a standard dynamically typed functional imperative language,
modeled on Scheme.
It has a call{-}by{-}value heap{-}based small{-}step semantics,
\raisebox{-2.3625bp}{\makebox[70.37968749999999bp][l]{\includegraphics[trim=2.4000000000000004 2.4000000000000004 2.4000000000000004 2.4000000000000004]{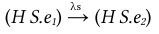}}}, where \raisebox{-2.617187499999999bp}{\makebox[6.459375bp][l]{\includegraphics[trim=2.4000000000000004 2.4000000000000004 2.4000000000000004 2.4000000000000004]{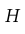}}}
represents the heap and \raisebox{-2.617187499999999bp}{\makebox[10.303125bp][l]{\includegraphics[trim=2.4000000000000004 2.4000000000000004 2.4000000000000004 2.4000000000000004]{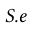}}} is represents a source
expression.\NoteBox{\NoteContent{We use a prefix followed by a dot (.) to distinguish terms in each
language{---}the prefix \emph{S} for source terms and the prefix \emph{A} for ANF
terms.}}
We omit the syntax and reduction rules for brevity.

The target language is essentially the same, but the syntax is restricted to
A{-}normal form: all computations \raisebox{-2.617187499999999bp}{\makebox[13.471875bp][l]{\includegraphics[trim=2.4000000000000004 2.4000000000000004 2.4000000000000004 2.4000000000000004]{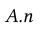}}} require values
\raisebox{-2.617187499999999bp}{\makebox[13.03125bp][l]{\includegraphics[trim=2.4000000000000004 2.4000000000000004 2.4000000000000004 2.4000000000000004]{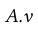}}} as operands; expressions \raisebox{-2.617187499999999bp}{\makebox[12.3515625bp][l]{\includegraphics[trim=2.4000000000000004 2.4000000000000004 2.4000000000000004 2.4000000000000004]{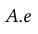}}}
cannot be nested and only explicitly compose and sequence intermediate
computations \raisebox{-2.617187499999999bp}{\makebox[13.471875bp][l]{\includegraphics[trim=2.4000000000000004 2.4000000000000004 2.4000000000000004 2.4000000000000004]{pict_4.pdf}}}.
The reduction relation, \raisebox{-2.3625bp}{\makebox[74.4765625bp][l]{\includegraphics[trim=2.4000000000000004 2.4000000000000004 2.4000000000000004 2.4000000000000004]{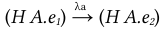}}},
does not require a control stack.

\begin{Figure}\begin{Centerfigure}\begin{FigureInside}\raisebox{-2.7265624999999956bp}{\makebox[178.6484375bp][l]{\includegraphics[trim=2.4000000000000004 2.4000000000000004 2.4000000000000004 2.4000000000000004]{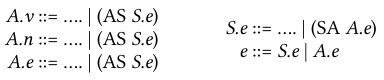}}}\end{FigureInside}\end{Centerfigure}

\Centertext{\Legend{\FigureTarget{\label{t:x28counter_x28x22figurex22_x22figx3aanfx2dmultix2dsynx22x29x29}\textsf{Fig.}~\textsf{1}. }{t:x28counter_x28x22figurex22_x22figx3aanfx2dmultix2dsynx22x29x29}\textsf{$\lambda$\textsuper{sa} Syntax (excerpts)}}}\end{Figure}

To develop a multi{-}language semantics, we embed syntactic terms from each
language into a single syntax, defined in \hyperref[t:x28counter_x28x22figurex22_x22figx3aanfx2dmultix2dsynx22x29x29]{Figure~\FigureRef{1}{t:x28counter_x28x22figurex22_x22figx3aanfx2dmultix2dsynx22x29x29}}.
We extend each meta{-}variable with boundary terms \raisebox{-2.617187499999999bp}{\makebox[31.7953125bp][l]{\includegraphics[trim=2.4000000000000004 2.4000000000000004 2.4000000000000004 2.4000000000000004]{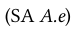}}}
({``}Source on the outside, ANF on the inside{''}) and \raisebox{-2.617187499999999bp}{\makebox[29.746875000000003bp][l]{\includegraphics[trim=2.4000000000000004 2.4000000000000004 2.4000000000000004 2.4000000000000004]{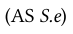}}}
({``}ANF on the outside, Source on the inside{''}).

\begin{Figure}\begin{Centerfigure}\begin{FigureInside}\raisebox{-3.2390624999999886bp}{\makebox[217.46328125bp][l]{\includegraphics[trim=2.4000000000000004 2.4000000000000004 2.4000000000000004 2.4000000000000004]{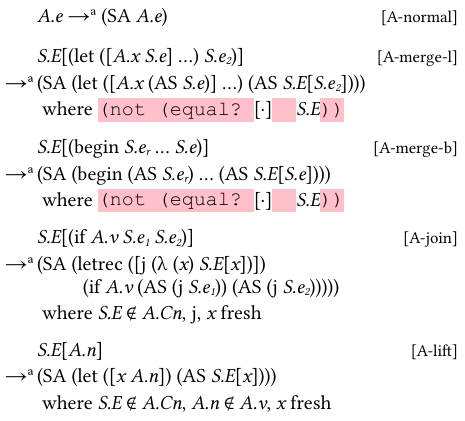}}}\end{FigureInside}\end{Centerfigure}

\Centertext{\Legend{\FigureTarget{\label{t:x28counter_x28x22figurex22_x22figx3aax2dredx22x29x29}\textsf{Fig.}~\textsf{2}. }{t:x28counter_x28x22figurex22_x22figx3aax2dredx22x29x29}\textsf{The A{-}reductions (excerpts)}}}\end{Figure}

The translation to ANF can be viewed as a reduction system in the
multi{-}language.
We define the A{-}reductions in \hyperref[t:x28counter_x28x22figurex22_x22figx3aax2dredx22x29x29]{Figure~\FigureRef{2}{t:x28counter_x28x22figurex22_x22figx3aax2dredx22x29x29}}.
These rules are essentially standard\Autobibref{~(\hyperref[t:x28autobib_x22Cormac_Flanaganx2c_Amr_Sabryx2c_Bruce_Fx2e_Dubax2c_and_Matthias_FelleisenThe_Essence_of_Compiling_with_ContinuationsIn_Procx2e_International_Conference_on_Programming_Language_Design_and_Implementation_x28PLDIx291993doix3a10x2e1145x2f155090x2e155113x22x29]{\AutobibLink{Flanagan et al\Sendabbrev{.}}} \hyperref[t:x28autobib_x22Cormac_Flanaganx2c_Amr_Sabryx2c_Bruce_Fx2e_Dubax2c_and_Matthias_FelleisenThe_Essence_of_Compiling_with_ContinuationsIn_Procx2e_International_Conference_on_Programming_Language_Design_and_Implementation_x28PLDIx291993doix3a10x2e1145x2f155090x2e155113x22x29]{\AutobibLink{1993}})}, but we modify them to
make boundary transitions explicit.
The A{-}reductions have the form \raisebox{-2.617187499999999bp}{\makebox[39.196875000000006bp][l]{\includegraphics[trim=2.4000000000000004 2.4000000000000004 2.4000000000000004 2.4000000000000004]{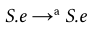}}}, reducing
source expressions in the multi{-}language.
Each A{-}reduction rewrites a source expression in a source evaluation context,
transforming the control stack into a data stack.
For example, the A{-}lift rule lifts a trivial computation, let{-}binding it and
providing the let{-}bound name (a value) in evaluation position, explicitly
sequencing the computation \raisebox{-2.617187499999999bp}{\makebox[13.471875bp][l]{\includegraphics[trim=2.4000000000000004 2.4000000000000004 2.4000000000000004 2.4000000000000004]{pict_4.pdf}}} with the evaluation context
\raisebox{-2.617187499999999bp}{\makebox[11.503125bp][l]{\includegraphics[trim=2.4000000000000004 2.4000000000000004 2.4000000000000004 2.4000000000000004]{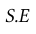}}}.
The side{-}conditions syntactically encode termination conditions, preventing
A{-}reductions of target redexes and in empty evaluation contexts.

\begin{Figure}\begin{Centerfigure}\begin{FigureInside}\raisebox{-0.7093750000000014bp}{\makebox[155.571875bp][l]{\includegraphics[trim=2.4000000000000004 2.4000000000000004 2.4000000000000004 2.4000000000000004]{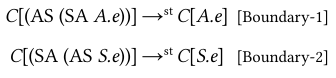}}}\end{FigureInside}\end{Centerfigure}

\Centertext{\Legend{\FigureTarget{\label{t:x28counter_x28x22figurex22_x22figx3aanfx2dboundaryx2dredx22x29x29}\textsf{Fig.}~\textsf{3}. }{t:x28counter_x28x22figurex22_x22figx3aanfx2dboundaryx2dredx22x29x29}\textsf{$\lambda$\textsuper{sa} Boundary Reductions}}}\end{Figure}

We supplement the multi{-}language A{-}reductions with the standard boundary
cancellation reductions, given in \hyperref[t:x28counter_x28x22figurex22_x22figx3aanfx2dboundaryx2dredx22x29x29]{Figure~\FigureRef{3}{t:x28counter_x28x22figurex22_x22figx3aanfx2dboundaryx2dredx22x29x29}}.
These apply under any multi{-}language context \raisebox{-2.617187499999999bp}{\makebox[5.910937499999999bp][l]{\includegraphics[trim=2.4000000000000004 2.4000000000000004 2.4000000000000004 2.4000000000000004]{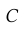}}}.

\begin{Figure}\begin{Centerfigure}\begin{FigureInside}\raisebox{-2.9625000000000012bp}{\makebox[254.3453125bp][l]{\includegraphics[trim=2.4000000000000004 2.4000000000000004 2.4000000000000004 2.4000000000000004]{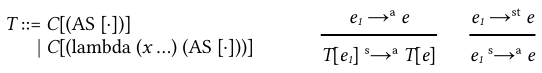}}}\end{FigureInside}\end{Centerfigure}

\Centertext{\Legend{\FigureTarget{\label{t:x28counter_x28x22figurex22_x22figx3aanfx2dtransx2dredx22x29x29}\textsf{Fig.}~\textsf{4}. }{t:x28counter_x28x22figurex22_x22figx3aanfx2dtransx2dredx22x29x29}\textsf{$\lambda$\textsuper{sa} Translation Reductions}}}\end{Figure}

In \hyperref[t:x28counter_x28x22figurex22_x22figx3aanfx2dtransx2dredx22x29x29]{Figure~\FigureRef{4}{t:x28counter_x28x22figurex22_x22figx3aanfx2dtransx2dredx22x29x29}} we define the translation reductions.
These extend the A{-}reductions to apply under any translation context
\raisebox{-2.617187499999999bp}{\makebox[5.221875000000001bp][l]{\includegraphics[trim=2.4000000000000004 2.4000000000000004 2.4000000000000004 2.4000000000000004]{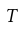}}}.
The construction of the translation context for ANF is a little unusual, but
the intuition is simple: a translation context identifies a pure source
expression under any context, including under a target/source boundary.
The context \raisebox{-2.617187499999999bp}{\makebox[21.834375bp][l]{\includegraphics[trim=2.4000000000000004 2.4000000000000004 2.4000000000000004 2.4000000000000004]{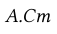}}} corresponds to an ANF context that can have
any expression in the hole.
In one step, the translation reductions can perform either one A{-}reduction
or one boundary cancellation.

From the translation reductions, we derive AOT compilation as
normalization with respect to translation reductions.
\Iidentity{\begin{definition}[ANF Compilation by Normalization]\Iidentity{\label{definition:anf:compilation:by:normalization}}\end{definition}}
\raisebox{-2.471875bp}{\makebox[116.0265625bp][l]{\includegraphics[trim=2.4000000000000004 2.4000000000000004 2.4000000000000004 2.4000000000000004]{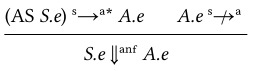}}}

\begin{Figure}\begin{Centerfigure}\begin{FigureInside}\raisebox{-0.3999999999999915bp}{\makebox[279.02890625000003bp][l]{\includegraphics[trim=2.4000000000000004 2.4000000000000004 2.4000000000000004 2.4000000000000004]{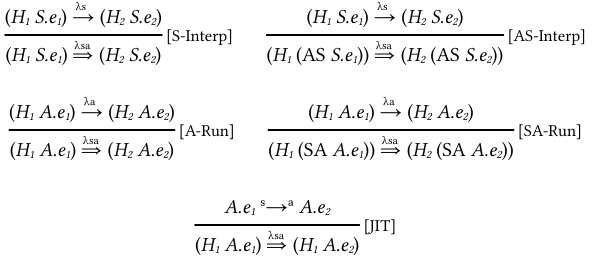}}}\end{FigureInside}\end{Centerfigure}

\Centertext{\Legend{\FigureTarget{\label{t:x28counter_x28x22figurex22_x22figx3aanfx2dmultix2dredx22x29x29}\textsf{Fig.}~\textsf{5}. }{t:x28counter_x28x22figurex22_x22figx3aanfx2dmultix2dredx22x29x29}\textsf{$\lambda$\textsuper{sa} Multi{-}language Reduction}}}\end{Figure}

Finally, we define the multi{-}language semantics in
\hyperref[t:x28counter_x28x22figurex22_x22figx3aanfx2dmultix2dredx22x29x29]{Figure~\FigureRef{5}{t:x28counter_x28x22figurex22_x22figx3aanfx2dmultix2dredx22x29x29}}.
This defines all possible transitions in the multi{-}language.
A term can either take a step in the source language, or a translation step, or
a step in the target language.
Multi{-}language reduction is indexed by a heap, \raisebox{-2.617187499999999bp}{\makebox[6.459375bp][l]{\includegraphics[trim=2.4000000000000004 2.4000000000000004 2.4000000000000004 2.4000000000000004]{pict_2.pdf}}}, which
is used by the source and target reductions but not the translation reductions.

Note that terms already in the heap are not translated, which corresponds to an
assumption that the language memory models are identical.
We could lift this restriction by adding multi{-}language boundaries to heap
values and extending translation reductions to apply in the heap.

The multi{-}language reduction allows reducing in the source, modeling
interpretation, or translating then reducing in the target, modeling
JIT compilation before continuing execution.
This does not model speculative optimization; equipping the multi{-}language with
assumption instructions as done by \Autobibref{\hyperref[t:x28autobib_x22Olivier_Flxfcckigerx2c_Gabriel_Schererx2c_Mingx2dHo_Yeex2c_Aviral_Goelx2c_Amal_Ahmedx2c_and_Jan_VitekCorrectness_of_speculative_optimizations_with_dynamic_deoptimizationProceedings_of_the_ACM_on_Programming_Languages_x28PACMPLx29_2x28POPLx29x2c_ppx2e_1x2dx2d282018doix3a10x2e1145x2f3158137x22x29]{\AutobibLink{Fl\"{u}ckiger et al\Sendabbrev{.}}}~(\hyperref[t:x28autobib_x22Olivier_Flxfcckigerx2c_Gabriel_Schererx2c_Mingx2dHo_Yeex2c_Aviral_Goelx2c_Amal_Ahmedx2c_and_Jan_VitekCorrectness_of_speculative_optimizations_with_dynamic_deoptimizationProceedings_of_the_ACM_on_Programming_Languages_x28PACMPLx29_2x28POPLx29x2c_ppx2e_1x2dx2d282018doix3a10x2e1145x2f3158137x22x29]{\AutobibLink{2018}})} might support
modeling this.

Standard meta{-}theoretic properites of reduction impliy standard compiler
correctness results.

Subject reduction of the multi{-}language semantics implies type{-}preservation of
the compiler.
This is simple for our present compiler, since the type system is simple, but
the theorems applies for more complex type systems.

\Iidentity{\begin{theorem}[Subject Reduction implies Type Preservation]\Iidentity{\label{theorem:thm:type-pres-type-pres}}If (\raisebox{-2.3625bp}{\makebox[33.55859375000001bp][l]{\includegraphics[trim=2.4000000000000004 2.4000000000000004 2.4000000000000004 2.4000000000000004]{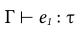}}} and \raisebox{-2.3625bp}{\makebox[37.22500000000001bp][l]{\includegraphics[trim=2.4000000000000004 2.4000000000000004 2.4000000000000004 2.4000000000000004]{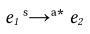}}}
implies \raisebox{-2.3625bp}{\makebox[33.55859375000001bp][l]{\includegraphics[trim=2.4000000000000004 2.4000000000000004 2.4000000000000004 2.4000000000000004]{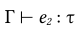}}}) then\Iidentity{\\}
(\raisebox{-2.617187499999999bp}{\makebox[50.32031250000001bp][l]{\includegraphics[trim=2.4000000000000004 2.4000000000000004 2.4000000000000004 2.4000000000000004]{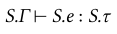}}} and \raisebox{-2.617187499999999bp}{\makebox[40.73437500000001bp][l]{\includegraphics[trim=2.4000000000000004 2.4000000000000004 2.4000000000000004 2.4000000000000004]{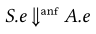}}} implies
${\exists}$\raisebox{-2.617187499999999bp}{\makebox[13.453125bp][l]{\includegraphics[trim=2.4000000000000004 2.4000000000000004 2.4000000000000004 2.4000000000000004]{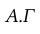}}},\raisebox{-2.617187499999999bp}{\makebox[12.435937500000001bp][l]{\includegraphics[trim=2.4000000000000004 2.4000000000000004 2.4000000000000004 2.4000000000000004]{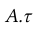}}}.
\raisebox{-2.617187499999999bp}{\makebox[56.465625bp][l]{\includegraphics[trim=2.4000000000000004 2.4000000000000004 2.4000000000000004 2.4000000000000004]{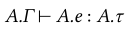}}}).\end{theorem}}

We derive compiler correctness from confluence.

\Iidentity{\begin{conjecture}[Confluence]\Iidentity{\label{conjecture:thm:anf:confluence}}If \raisebox{-2.3625bp}{\makebox[58.870312500000004bp][l]{\includegraphics[trim=2.4000000000000004 2.4000000000000004 2.4000000000000004 2.4000000000000004]{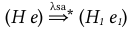}}} and \Iidentity{\\}
\raisebox{-2.3625bp}{\makebox[58.870312500000004bp][l]{\includegraphics[trim=2.4000000000000004 2.4000000000000004 2.4000000000000004 2.4000000000000004]{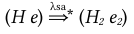}}} then
\raisebox{-2.3625bp}{\makebox[63.840624999999996bp][l]{\includegraphics[trim=2.4000000000000004 2.4000000000000004 2.4000000000000004 2.4000000000000004]{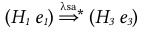}}} and
\raisebox{-2.3625bp}{\makebox[63.840624999999996bp][l]{\includegraphics[trim=2.4000000000000004 2.4000000000000004 2.4000000000000004 2.4000000000000004]{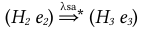}}}\end{conjecture}}

Note the multi{-}language semantics can reduce open terms, so confluence implies
correctness of both the AOT and the JIT compiler.
As an example, whole{-}program correctness is a trivial corollary of confluence.

\Iidentity{\begin{corollary}[Whole-Program Correctness]\Iidentity{\label{corollary:thm:anf:correct}}\Iidentity{\mbox{}\\}
If
\raisebox{-2.617187499999999bp}{\makebox[67.2875bp][l]{\includegraphics[trim=2.4000000000000004 2.4000000000000004 2.4000000000000004 2.4000000000000004]{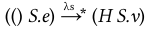}}} and
\raisebox{-2.617187499999999bp}{\makebox[40.73437500000001bp][l]{\includegraphics[trim=2.4000000000000004 2.4000000000000004 2.4000000000000004 2.4000000000000004]{pict_25.pdf}}} then
\raisebox{-2.617187499999999bp}{\makebox[71.38437499999999bp][l]{\includegraphics[trim=2.4000000000000004 2.4000000000000004 2.4000000000000004 2.4000000000000004]{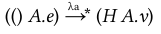}}} such that
\raisebox{-2.617187499999999bp}{\makebox[13.03125bp][l]{\includegraphics[trim=2.4000000000000004 2.4000000000000004 2.4000000000000004 2.4000000000000004]{pict_5.pdf}}} is equal to
\raisebox{-2.617187499999999bp}{\makebox[10.9828125bp][l]{\includegraphics[trim=2.4000000000000004 2.4000000000000004 2.4000000000000004 2.4000000000000004]{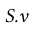}}}.\end{corollary}}

Multi{-}language semantics provide a strong attacker model through contextual
equivalence.
A context \raisebox{-2.617187499999999bp}{\makebox[5.910937499999999bp][l]{\includegraphics[trim=2.4000000000000004 2.4000000000000004 2.4000000000000004 2.4000000000000004]{pict_15.pdf}}} models an attacker that can provide
either source or target code or data as input and observe the result.
Contextual equivalence is extended to relate reduction configurations, not
just terms, to enable the definition to apply to the JIT model.

\Iidentity{\begin{definition}[Contextual Equivalence]\Iidentity{\label{definition:contextual:equivalence}}\raisebox{-2.3625bp}{\makebox[23.396875bp][l]{\includegraphics[trim=2.4000000000000004 2.4000000000000004 2.4000000000000004 2.4000000000000004]{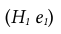}}} \Iidentity{$\mathrel{\approx}$} \raisebox{-2.3625bp}{\makebox[23.396875bp][l]{\includegraphics[trim=2.4000000000000004 2.4000000000000004 2.4000000000000004 2.4000000000000004]{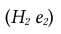}}}
if for all multi-language contexts \raisebox{-2.617187499999999bp}{\makebox[5.910937499999999bp][l]{\includegraphics[trim=2.4000000000000004 2.4000000000000004 2.4000000000000004 2.4000000000000004]{pict_15.pdf}}}, \raisebox{-2.3625bp}{\makebox[36.142187500000006bp][l]{\includegraphics[trim=2.4000000000000004 2.4000000000000004 2.4000000000000004 2.4000000000000004]{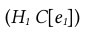}}} and \raisebox{-2.3625bp}{\makebox[36.142187500000006bp][l]{\includegraphics[trim=2.4000000000000004 2.4000000000000004 2.4000000000000004 2.4000000000000004]{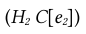}}} co-terminate
in \raisebox{-2.9390624999999986bp}{\makebox[6.84375bp][l]{\includegraphics[trim=2.4000000000000004 2.4000000000000004 2.4000000000000004 2.4000000000000004]{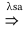}}}.\end{definition}}

We define secure compilation of both the AOT and JIT models as full abstraction:
contextual equivalence is preserved and reflected through multi{-}language
reduction.

\Iidentity{\begin{theorem}[Full Abstraction (multi-language)]\Iidentity{\label{theorem:full:abstraction:(multi-language)}}Suppose
\raisebox{-2.3625bp}{\makebox[66.67343749999999bp][l]{\includegraphics[trim=2.4000000000000004 2.4000000000000004 2.4000000000000004 2.4000000000000004]{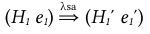}}} and
\raisebox{-2.3625bp}{\makebox[66.67343749999999bp][l]{\includegraphics[trim=2.4000000000000004 2.4000000000000004 2.4000000000000004 2.4000000000000004]{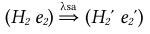}}}.
\Iidentity{\\}Then
\raisebox{-2.3625bp}{\makebox[23.396875bp][l]{\includegraphics[trim=2.4000000000000004 2.4000000000000004 2.4000000000000004 2.4000000000000004]{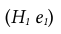}}} \Iidentity{$\mathrel{\approx}$} \raisebox{-2.3625bp}{\makebox[23.396875bp][l]{\includegraphics[trim=2.4000000000000004 2.4000000000000004 2.4000000000000004 2.4000000000000004]{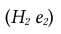}}}
if and only if
\raisebox{-2.3625bp}{\makebox[28.168750000000003bp][l]{\includegraphics[trim=2.4000000000000004 2.4000000000000004 2.4000000000000004 2.4000000000000004]{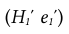}}} \Iidentity{$\mathrel{\approx}$}
\raisebox{-2.3625bp}{\makebox[28.168750000000003bp][l]{\includegraphics[trim=2.4000000000000004 2.4000000000000004 2.4000000000000004 2.4000000000000004]{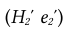}}}.\end{theorem}}

The normally easy part of full abstraction, within the multi{-}language, is now a
direct consequence of confluence, since both compilation and contextual
equivalence are defined by multi{-}language reduction.
The hard part, showing any multi{-}language context (attacker) is emulated by a
source context, remains.

\sectionNewpage

\Ssectionstarx{Bibliography}{Bibliography}\label{t:x28part_x22docx2dbibliographyx22x29}

\begin{AutoBibliography}\begin{SingleColumn}\Autobibtarget{\label{t:x28autobib_x22Amal_AhmedVerified_Compilers_for_a_Multix2dlanguage_WorldIn_Procx2e_Summit_oN_Advances_in_Programming_Languages_x28SNAPLx292015doix3a10x2e4230x2fLIPIcsx2eSNAPLx2e2015x2e15x22x29}\Autobibentry{Amal Ahmed. Verified Compilers for a Multi-language World. In \textit{Proc. Summit oN Advances in Programming Languages (SNAPL)}, 2015. \pseudodoi{doi:\href{https://doi.org/10.4230/LIPIcs.SNAPL.2015.15}{10{\hbox{\texttt{.}}}4230/LIPIcs{\hbox{\texttt{.}}}SNAPL{\hbox{\texttt{.}}}2015{\hbox{\texttt{.}}}15}}}}

\Autobibtarget{\label{t:x28autobib_x22Amal_Ahmed_and_Matthias_BlumeAn_Equivalencex2dPreserving_CPS_Translation_via_Multix2dLanguage_SemanticsIn_Procx2e_International_Conference_on_Functional_Programming_x28ICFPx292011doix3a10x2e1145x2f2034773x2e2034830x22x29}\Autobibentry{Amal Ahmed and Matthias Blume. An Equivalence-Preserving CPS Translation via Multi-Language Semantics. In \textit{Proc. International Conference on Functional Programming (ICFP)}, 2011. \pseudodoi{doi:\href{https://doi.org/10.1145/2034773.2034830}{10{\hbox{\texttt{.}}}1145/2034773{\hbox{\texttt{.}}}2034830}}}}

\Autobibtarget{\label{t:x28autobib_x22Cormac_Flanaganx2c_Amr_Sabryx2c_Bruce_Fx2e_Dubax2c_and_Matthias_FelleisenThe_Essence_of_Compiling_with_ContinuationsIn_Procx2e_International_Conference_on_Programming_Language_Design_and_Implementation_x28PLDIx291993doix3a10x2e1145x2f155090x2e155113x22x29}\Autobibentry{Cormac Flanagan, Amr Sabry, Bruce F. Duba, and Matthias Felleisen. The Essence of Compiling with Continuations. In \textit{Proc. International Conference on Programming Language Design and Implementation (PLDI)}, 1993. \pseudodoi{doi:\href{https://doi.org/10.1145/155090.155113}{10{\hbox{\texttt{.}}}1145/155090{\hbox{\texttt{.}}}155113}}}}

\Autobibtarget{\label{t:x28autobib_x22Olivier_Flxfcckigerx2c_Gabriel_Schererx2c_Mingx2dHo_Yeex2c_Aviral_Goelx2c_Amal_Ahmedx2c_and_Jan_VitekCorrectness_of_speculative_optimizations_with_dynamic_deoptimizationProceedings_of_the_ACM_on_Programming_Languages_x28PACMPLx29_2x28POPLx29x2c_ppx2e_1x2dx2d282018doix3a10x2e1145x2f3158137x22x29}\Autobibentry{Olivier Fl\"{u}ckiger, Gabriel Scherer, Ming{-}Ho Yee, Aviral Goel, Amal Ahmed, and Jan Vitek. Correctness of speculative optimizations with dynamic deoptimization. \textit{Proceedings of the ACM on Programming Languages (PACMPL)} 2(POPL), pp. 1{--}28, 2018. \pseudodoi{doi:\href{https://doi.org/10.1145/3158137}{10{\hbox{\texttt{.}}}1145/3158137}}}}

\Autobibtarget{\label{t:x28autobib_x22Robert_Bruce_Matthews_Jacob_And_FindlerOperational_Semantics_for_Multix2dlanguage_ProgramsIn_Procx2e_Symposium_on_Principles_of_Programming_Languages_x28POPLx292007doix3a10x2e1145x2f1190216x2e1190220x22x29}\Autobibentry{Robert Bruce Matthews Jacob And Findler. Operational Semantics for Multi-language Programs. In \textit{Proc. Symposium on Principles of Programming Languages (POPL)}, 2007. \pseudodoi{doi:\href{https://doi.org/10.1145/1190216.1190220}{10{\hbox{\texttt{.}}}1145/1190216{\hbox{\texttt{.}}}1190220}}}}

\Autobibtarget{\label{t:x28autobib_x22Max_Sx2e_Newx2c_William_Jx2e_Bowmanx2c_and_Amal_AhmedFully_Abstract_Compilation_via_Universal_EmbeddingIn_Procx2e_International_Conference_on_Functional_Programming_x28ICFPx292016doix3a10x2e1145x2f2951913x2e2951941x22x29}\Autobibentry{Max S. New, William J. Bowman, and Amal Ahmed. Fully Abstract Compilation via Universal Embedding. In \textit{Proc. International Conference on Functional Programming (ICFP)}, 2016. \pseudodoi{doi:\href{https://doi.org/10.1145/2951913.2951941}{10{\hbox{\texttt{.}}}1145/2951913{\hbox{\texttt{.}}}2951941}}}}

\Autobibtarget{\label{t:x28autobib_x22Daniel_Patterson_and_Amal_AhmedLinking_Types_for_Multix2dLanguage_Softwarex3a_Have_Your_Cake_and_Eat_It_TooIn_Procx2e_Summit_oN_Advances_in_Programming_Languages_x28SNAPLx292017doix3a10x2e4230x2fLIPIcsx2eSNAPLx2e2017x2e12x22x29}\Autobibentry{Daniel Patterson and Amal Ahmed. Linking Types for Multi-Language Software: Have Your Cake and Eat It Too. In \textit{Proc. Summit oN Advances in Programming Languages (SNAPL)}, 2017. \pseudodoi{doi:\href{https://doi.org/10.4230/LIPIcs.SNAPL.2017.12}{10{\hbox{\texttt{.}}}4230/LIPIcs{\hbox{\texttt{.}}}SNAPL{\hbox{\texttt{.}}}2017{\hbox{\texttt{.}}}12}}}}

\Autobibtarget{\label{t:x28autobib_x22James_Tx2e_Perconti_and_Amal_AhmedVerifying_an_Open_Compiler_Using_Multix2dlanguage_SemanticsIn_Procx2e_European_Symposium_on_Programming_x28ESOPx292014doix3a10x2e1007x2f978x2d3x2d642x2d54833x2d8x5f8x22x29}\Autobibentry{James T. Perconti and Amal Ahmed. Verifying an Open Compiler Using Multi-language Semantics. In \textit{Proc. European Symposium on Programming (ESOP)}, 2014. \pseudodoi{doi:\href{https://doi.org/10.1007/978-3-642-54833-8_8}{10{\hbox{\texttt{.}}}1007/978{-}3{-}642{-}54833{-}8{\char`\_}8}}}}\end{SingleColumn}\end{AutoBibliography}

\postDoc
\end{document}